\documentclass[twocol]{ametsocV6.1}

\title{The ARP-GEM1 Global Atmosphere Model. Part II: Multiscale Evaluation up to 6 km.} 

\authors{Olivier GEOFFROY,\aff{a}\correspondingauthor{Olivier Geoffroy, olivier.geoffroy@meteo.fr} and David SAINT-MARTIN\aff{a}}
\affiliation{\aff{a}{CNRM, Université de Toulouse, Météo-France, CNRS, Toulouse, France}}

\abstract{This is the second part of a series of two articles focused on the development and evaluation of the ARP-GEM1 global atmosphere model. The first paper introduced the model's new physics and speedup improvements. In this second part, we evaluate ARP-GEM1 through a set of 30-year prescribed sea-surface-temperature simulations at 55, 25, 12.6, and 6.3 km resolutions. The model demonstrates reliability in representing global climate metrics, comparable to the top-performing CMIP6 models, while maintaining high computational efficiency at all resolutions. These simulations further demonstrate the feasibility of O(10)-km climate simulations, and highlight their added value, particularly for capturing phenomena such as cyclones. Ultimately, these exploratory simulations should be considered an intermediate step toward the development and tuning of even higher-resolution, convection-permitting kilometer-scale configurations.}
\begin{document}

\maketitle

\section{Introduction}

In Part I (Saint-Martin and Geoffroy 2024, hereafter Part I), the first version of the ARP-GEM atmospheric model (hereafter ARP-GEM1) is described in detail, along with a thorough analysis of its computational performance. ARP-GEM is a computationally efficient variant of the ARPEGE/IFS atmospheric model, incorporating an suite of state-of-the-art physical parameterizations. 
This model is approximately 15 times faster than the atmospheric component of CNRM-CM6-1 \citep{voldoire-2019} and CNRM-CM6-1-HR \citep{saintmartin-2021}, both of which participated in CMIP6.
The significant speedup results from the combined use of several acceleration strategies, including code optimization (general, mass conservation, I/O), the use of coarsened grids for radiative and surface computations, bit reduction for floating-point operations, reduced vertical levels, and the adoption of octahedral and cubic grids.

Concurrently, the ARP-GEM1 model incorporates simplifications and improvements in its physical processes compared to the CMIP6 version, including the use of a new deep convection scheme \citep{tiedtke-1989,bechtold-2008, bechtold-2014} with modifications to its triggering, a new shallow convection scheme, an updated cloud scheme \citep{smith-1990}, a new radiation scheme \citep{hogan-2015}, and additional developments in turbulence and microphysics.

Over the past decades, climate model errors have been slightly reduced, mainly due to refinements in atmospheric parameterizations and in the model tuning procedure \citep[e.g.][]{schneider-2024}. However, persistent biases remain, resulting from the use of approximate subgrid-scale representations of moist atmospheric processes \citep[e.g.][]{stevens-2013}. These errors can still be as large as the climate change signals \citep{palmer-2019}.

Refining the horizontal grid spacing is expected to reduce these errors. At resolutions around 25 km, general circulation models are better able to capture extreme precipitation events and the overall structure of tropical cyclones, although finer details require even higher resolutions \citep[e.g.,][]{wehner-2014}. However, phenomena related to the representation of convective motions remain deficient in these models, including the representation of the Inter-Tropical Convergence Zone and the Madden-Julian Oscillation (MJO) \citep[e.g.,][]{suematsu-2022}. To overcome the limitations of parameterized deep convection, horizontal grid spacing must be refined to at least a few kilometers to explicitly represent convective motions.

Kilometer-scale global convection-permitting models \citep[e.g.,][]{satoh-2005,satoh-2008} can improve the representation of phenomena such as extreme precipitation \citep[e.g.,][]{stevens-2020}, the propagation of convective storms \citep[e.g.,][]{marsham-2013}, the diurnal cycle \citep[e.g.,][]{hohenegger-2009, yashiro-2016}, as well as providing more detailed simulations of cyclones \citep[e.g.,][]{judt-2021}. However, they currently show mixed results for phenomena such as the MJO and the climatology of tropical precipitation \citep{wedi-2020,takasuka-2024}. Global convection-permitting simulations are currently limited to short durations, with typical durations of a few months in the latest DYAMOND intercomparison exercise \citep{stevens-2019}. Their high computational cost prevents systematic model calibration and the testing of new parameterizations \citep{schneider-2024}.

Intermediate resolutions can help address these issues. \cite{takasuka-2024} emphasize the importance of calibrating their 14-km resolution configuration to achieve reliable kilometer-scale climate simulations. \cite{wedi-2020} shows that their 9-km simulation with parameterized deep convection yields comparable results to their 1.4 km simulation. Moreover, analyzing a range of resolutions can help document the added value of increased resolutions and identify those that achieve the optimal balance between accuracy and efficiency \citep{hohenegger-2020}.

In this study, we present results from a set of 30-year prescribed sea-surface-temperature simulations at 55, 25, 12.6, and 6.3 km resolutions using ARP-GEM1. A particular effort has been made in these experiments to provide a 'multiscale' tuning, i.e., to calibrate and optimize the model settings consistently at the different resolutions. The analysis is two-fold: first, they provide a multiscale evaluation of the main climate features of the ARP-GEM1 model in \textit{amip} mode. Second, they offer insights into the added value and potential emerging problems associated with increasing horizontal resolution from 55 to 6 km. This serves as an intermediate step towards the setup of pluriannual convection-permitting experiments with the ARP-GEM model.

The article is organized as follows. Section \ref{sec:simus} describes the experiments and details the calibration strategy. Section \ref{sec:evaluation} evaluates the model's main climate characteristics.

\section{Simulation suite at 55 to 6 km resolution}
\label{sec:simus}

\subsection{Resolution configurations and computational costs}
\label{res:costs}

\begin{table*}[h]
\caption{Configuration details and computational performance for the four simulations. The grid o$N_g$ refers to a octahedral reduced Gaussian grid with $N_g$ Gaussian latitudes and 2$N_g$ longitudes along equatorial Gaussian latitudes. The coarsening factor refers to the ratio between the grid-point model resolution and the radiative grid resolution. SYPD refers to as Simulated Years Per Day.}
\begin{center}
\begin{tabular}{lcccccccc}
\topline
Configuration & Grid Point & Spectral & Time & Rad. \& Surf. & Coarsening & Radiation & CPU & SYPD \\
Name & Resolution (km) & Truncation & Step (s) & Resolution & Facto & Timestep (s) & Cores &  \\
\midline
ARP-GEM1-55km & o360 (55 km)  & 179  & 900 & o128 (156 km)  & 2.8 & 7200 & 9x128 & 77.0 \\
ARP-GEM1-25km & o782 (25 km)  & 390  & 900 & o244  (82 km) & 3.2 & 7200 & 9x128 & 15.8 \\
ARP-GEM1-12km & o1564 (12.6 km)& 781  & 600 & o488  (41 km) & 3.2 & 3600 & 18x128 & 7.0 \\
ARP-GEM1-6km  & o3048 (6.3 km)  & 1523 & 300 & o958  (21 km) & 3.2 & 3600 & 30x128 & 1.6 \\
\botline
\end{tabular}
\label{tab:configurations}
\end{center}
\end{table*}

Thirty-year prescribed sea-surface-temperature simulations using version 1 of the ARP-GEM model are presented here. The forcings are similar to the CMIP6 \textit{amip} experiment and cover the period 1985-2014. Paired 30-year simulations have also been carried out with the \textit{amip-future4K} forcings, but these are not analyzed in this article.
We ran a suite of four configurations -- 55 km, 25 km, 12.6 km, and 6.3 km resolutions. Configurations are referred to as ARP-GEM1-55km, ARP-GEM1-25km, ARP-GEM1-12km, ARP-GEM1-6km, respectively.

Table \ref{tab:configurations} highlights the main differences between the simulation features. Radiation calculations are performed every two hours for the 55-km and 25-km resolutions, and every hour for the 12-km and 6-km resolutions. Differences in physical parameters are discussed in the next subsection. Using nine nodes (1152 CPU cores), the 55-km simulation requires 19 minutes of machine time to simulate one year, corresponding to approximately 77 simulated years per day (SYPD). With 30 nodes, representing 3\% of CNRM's computational resources, the 6-km configuration achieves a SYPD of 1.6.

\begin{figure}
\centerline{\includegraphics[width=19pc]{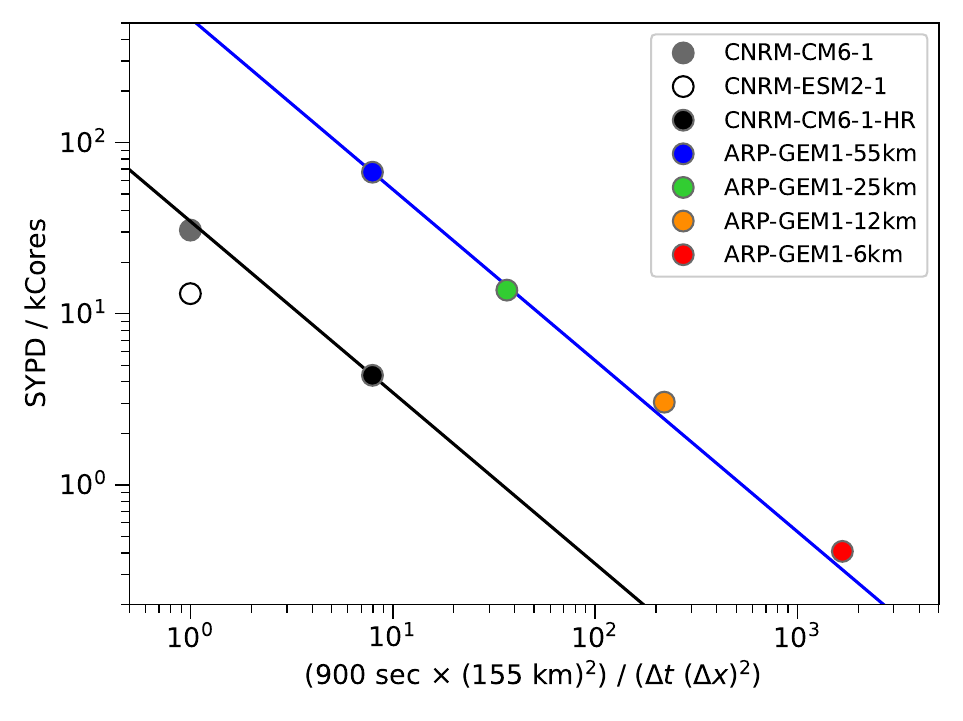}}
\caption{Simulated years per day (SYPD) in a 24-h period normalized by the number of kilo-cores used as a function of normalized timestep and grid spacing for the CNRM CMIP6 models : CNRM-CM6-1 (black), CNRM-ESM2-1 (white), CNRM-CM6-1-HR (gray) and for the ARP-GEM1 configurations : at 55-km (blue), 25-km (green), 12-km (orange) and 6-km (red) resolutions. The x-axis is chosen as the value for the CNRM-CM6-1 model is equal to 1. The lines represent computational costs estimated from an idealized scaling of the 55-km simulations (see text).}
\label{fig:sypd}
\end{figure}

Figure \ref{fig:sypd} compares the computational cost of each model configuration. The SYPD, normalized by the number of cores, is plotted against the normalized product of spatial and temporal resolution. The lines represent idealized scaling, illustrating how computational cost changes based on linear scaling with the square of horizontal resolution and the timestep. For instance, with the same computational resources, a 12.5-km resolution with a 300-second timestep is expected to to be 8 times slower than a 25-km resolution with a 600-second timestep. All four data points align closely with this scaling line, indicating that the efficiency of the ARP-GEM model, documented in Part I, is preserved in resolutions up to 6 km. 

In comparison, results from similar \textit{amip} experiments within the CMIP6 versions of the CNRM climate model are also presented. The CNRM-CM6-1 configuration (approximately 155-km resolution with a 900-second timestep) and the CNRM-CM6.1-HR (approximately 55-km resolution with the same timestep) exhibit consistent scaling. The CNRM-ESM2.1  runs at roughly half the speed of its counterparts. Despite this, all three CMIP6 configurations produce similar results in terms of physical model performance \citep[e.g.,][]{seferian-2019}. The computational cost of the ARP-GEM1-25km \textit{amip} experiment is comparable to that of the CNRM-ESM2.1 \textit{amip} experiment (155-km resolution).

\subsection{Tuning strategy}
\label{sec:tuning}

Model development and associated calibrations have been performed using 1D configuration and 55-km simulations. Most of the final tuning has been carried out using the 25-km resolution configuration, with a few simulations at 12 km to help finalize the values of some parameters, though these changes did not significantly alter the model state. Note that the 6-km and 12-km simulations are relatively inexpensive, making further tuning tests feasible, particularly at 12 km, which allows for an extensive number of simulations.

A minimal set of parameters was selected to adjust for changes related to spatial resolution and associated timestep. The convective closure time scale parameter $k_{\mathrm{cv}}$ (see Part I) was adjusted a priori in the 6-km simulation based on established dependencies, as outlined in \citet{ecmwf-2019}. \citet{ecmwf-2019} hypothesizes that the convective closure time scale increases for resolutions above 8 km and decreases logarithmically for resolutions below 8 km. In our experiments, we applied the scaling factor from their formulation only for resolutions finer than 8 km, not considering the dependency for lower resolutions. It is worth noting that using a constant value across all simulations might have been a more effective approach for evaluating sensitivity to resolution.

The final tuning was conducted using a minimal set of three parameters : the low level cloud critical relative humidity $RH_{\mathrm{c},\mathrm{low}}$, the ice autoconversion rate scaling factor $k_{\mathrm{au},i}$ and the SW inhomogeneity factor IF$_{\mathrm{sw}}$, as detailed below, with a particular emphasis on the radiation budget and cloud cover. First, the $RH_{\mathrm{c},\mathrm{low}}$ parameter was modified to address changes in low cloud fraction, which decreased (by a few percent) as spatial resolution increased and timestep decreased. The final radiative adjustments were made by modifying only $k_{\mathrm{au},i}$ and IF$_{\mathrm{sw}}$. 

Shortwave radiation (SW) was adjusted directly by modifying cloud optical depth and longwave radiation (LW) was adjusted indirectly through changes in high cloud amounts. The high cloud amount was modified by altering the ice autoconversion rate scaling factor $k_{\mathrm{au},i}$, as LW high cloud radiative effect is influenced by large-scale microphysics. Note that this effect is not highly sensitive to the cloud scheme parameters. Moreover, the impact of LW inhomogeneity factor is minimal, so this parameter is not used for LW. For simplicity, the ice and liquid autoconversion parameters assigned identical values and treated as a single parameter, although more precise tuning could differentiate them.

The use of the inhomogeneity factor parameter can be justified as relatively physical \citep{oreopoulos-2005}. Nevertheless, it primarily functions as a final tuning mechanism, providing a coarse adjustment for radiation-cloud interactions. Due to compensating errors in cloud representation and the substantial uncertainty associated with the inhomogeneity factor, both ice and liquid inhomogeneity factors are treated as a single parameter. The ice autoconversion parameter (and, to a lesser extent, the liquid autoconversion parameter) is used to adjust LW radiation by influencing high-level cloud amounts. The inhomogeneity factor parameter is ultimately applied to adjust SW radiation, which is also influenced by the autoconversion rates.

Table \ref{tab:tuning} summarizes the parameter values used to adjust for changes in radiation and cloud cover with resolution. It also presents the associated global annual mean values of key climate variables used for tuning, including top-of-the-atmosphere radiation and total cloud cover, which play a critical role in the radiation budget. All simulations are in close agreement with observations, except for the LW CRE, which is lower than observational estimates. An underestimation is expected due to variations in estimation methodologies and the impact of water vapor on clear-sky LW fluxes \citep[e.g.][]{loeb-2020}. Part of the difference may also be attributed to an overly moist troposphere in the model (Section \ref{sec:evaluation}\ref{sub:ta_hus}), which is consistent with a low LW CRE.
The following sections provide an evaluation of all four simulations.

\section{Evaluation of the different resolutions}
\label{sec:evaluation}

\subsection{Mean state errors and comparison with CMIP6}
\label{res:rmse}

\begin{table*}
\caption{Tuning parameters and global mean radiation budget.}
\begin{center}
\begin{tabular}{lccccccccc}
\topline
& IF$_{\mathrm{sw}}$ & $k_{\mathrm{au},i/l}$ & $RH_{\mathrm{c},\mathrm{low}}$ & $k_{\mathrm{cv}}$ & \textbf{rlut} & \textbf{rst} & \textbf{rstcre} & \textbf{rltcre} & \textbf{clt} \\
&  & (s$^{-1}$) &  &  & (W m$^{-2}$) & (W m$^{-2}$) & (W m$^{-2}$) & (W m$^{-2}$) & (\%) \\
\midline
ARP-GEM1-55km & 0.780 & 7.0e-4 & 0.94 & 1.35 & 239.6 & 240.1 & -45.6 & 22.7 & 63.4 \\
ARP-GEM1-25km & 0.775 & 6.8e-4 & 0.94 & 1.35 & 239.5 & 240.3 & -45.2 & 22.7 & 63.7 \\
ARP-GEM1-12km & 0.865 & 6.2e-4 & 0.93 & 1.35 & 239.6 & 240.5 & -46.0 & 22.2 & 63.7 \\
ARP-GEM1-6km  & 0.945 & 4.0e-4 & 0.92 & 1.50 & 239.5 & 240.1 & -46.3 & 22.1 & 64.4 \\
Observations  &  &  &  &  & 239.7 & 240.5 & -47.1 & 26.0 & 66.5 \\
\botline
\end{tabular}
\label{tab:tuning}
\end{center}
\end{table*}

We first evaluate the performance of the ARP-GEM1 model configurations in simulating key climate variables, including top-of-the-atmosphere longwave and shortwave radiation, precipitation, surface air temperature, and total cloud cover. All these variables are energetically significant. Precipitation and surface air temperature also serve as major climate indicators, while cloud cover plays a significant role in the radiative budget. 

\begin{figure*}
\centerline{\includegraphics[width=33pc]{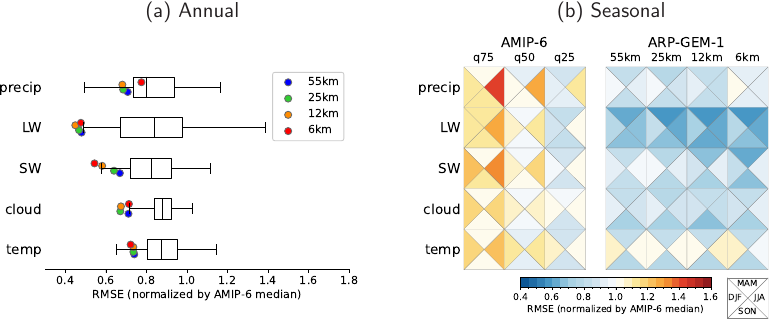}}
\caption{Annual and seasonal normalized root-mean square errors (RMSEs) in the climatology of precipitation (precip), top-of-atmosphere longwave (LW) and net shortwave (SW) radiation, total cloud cover (cloud) and surface air temperature (temp). RMSE is normalized by the median across 38 CMIP-6 models (see Appendix) for each field and across all seasons. (a) The boxplot shows the distribution of the annual RMSEs for the 38 CMIP-6 models in comparison with the annual RMSEs for ARP-GEM1 experiments : 55 (blue), 25 (green), 12 (orange), and 6-km (red) resolution. (b) The seasonal values of RMSEs for ARP-GEM1 experiments are compared with the first (q25) and third (q75) quantiles, along with the median value (q50).}
\label{fig:rmse}
\end{figure*}

We compare the root mean square error (RMSE) for these variables across the various resolutions of the ARP-GEM1 model and a large ensemble of 38 CMIP6 model versions (see Appendix). Climatologies of CMIP6 models are computed using the \textit{amip} experiment. RMSEs are assessed against climatologies : from CERES-EBAF \citep{loeb-2009} over the period 2001-2014 for SW and LW radiation, from the Multi‐Source Weighted‐Ensemble Precipitation (MSWEP) dataset, version 1.2 \citep{beck-2017} over the period 1985-2014 for the precipitation, from the CALIPSO-GOCCP product \citep{chepfer-2010} over the period 2007-2012 for the cloud cover, and from the BEST monthly data set \citep{rohde-2013} over the period 1985-2014 for the near-surface air temperature. All data are conservatively remapped to a common 2.5$^{\circ}$ regular grid.

Figure \ref{fig:rmse}a shows the normalized RMSE for the annual mean climatology, alongside the RMSE distribution of the CMIP6 ensemble, represented as a boxplot. Figure \ref{fig:rmse}b presents the RMSE of the seasonal mean climatologies for each ARP-GEM1 configuration, compared with the 75th, 50th, and 25th percentiles of the CMIP6 RMSE distributions.

\begin{figure*}
\centerline{\includegraphics[width=33pc]{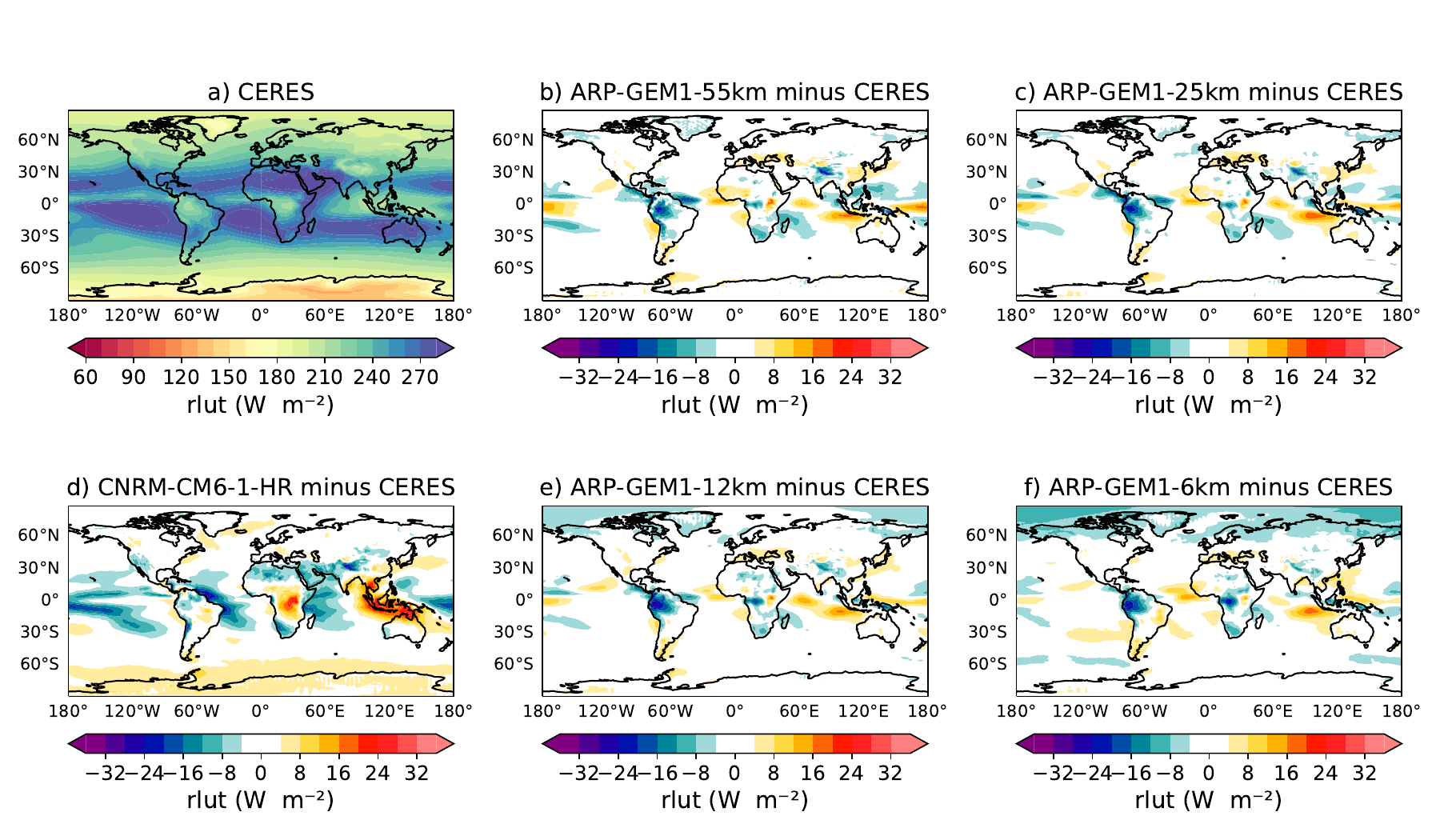}}
\caption{Annual mean of the top-of-the-atmosphere longwave radiation (period: 2001-2014) for : (a) CERES-EBAF data and for (b) ARP-GEM1-55km, (c) ARP-GEM1-25km, (d) CNRM-CM6-1-HR, (e) ARP-GEM1-12km, and (f) ARP-GEM1-6km biases.}
\label{fig:map_rlut}
\end{figure*}

\begin{figure*}
\centerline{\includegraphics[width=33pc]{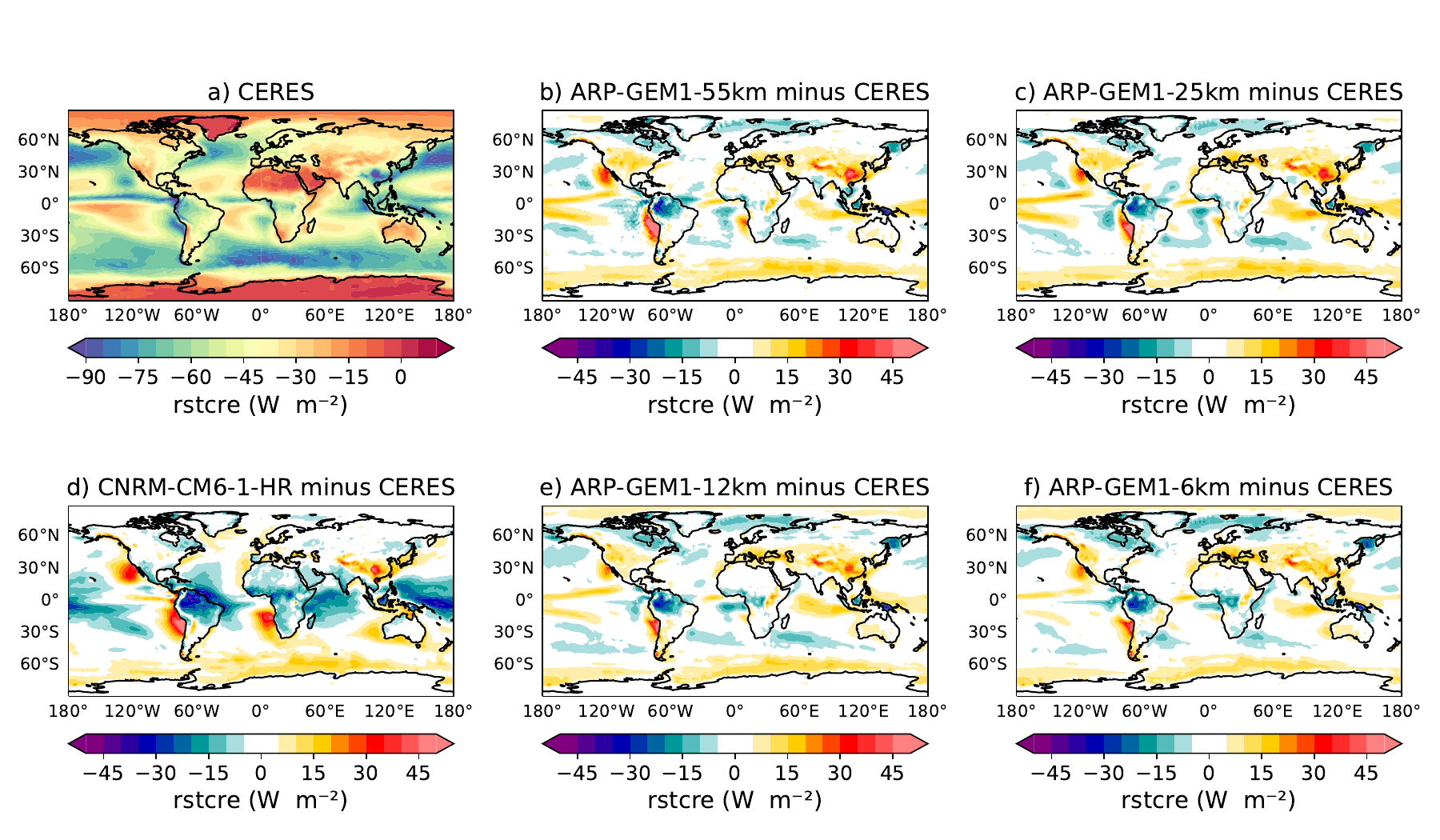}}
\caption{Same as Fig. \ref{fig:map_rlut} for top-of-the-atmosphere shortwave cloud radiative effect.}
\label{fig:map_rstcre}
\end{figure*}

The results show that the ARP-GEM1 model performs well at all resolutions, with its RMSE values often within or better than the 25th percentile of the CMIP6 ensemble distribution for both annual and seasonnal climatologies. Specifically, ARP-GEM1 outperforms three-quarters of CMIP6 models in annual climatologies and is close to the best CMIP6 model in several metrics. For instance, ARP-GEM1 is notably better than the best CMIP6 models in simulating cloud cover and show minimal errors in top-of-the-atmosphere radiation, especially longwave. It also performs well in precipitation and surface air temperature.

On first inspection, model resolution appears comparable across configurations, with minimal variability relative to the CMIP6 spread. This indicates that the minimal tuning strategy is effective.
The next section will evaluate the performance of the model across these different resolutions, considering the impact of model resolution on various aspects.

\subsection{Clouds and Radiation}
\label{sec:clouds}

\begin{figure*}[h]
\centerline{\includegraphics[width=33pc]{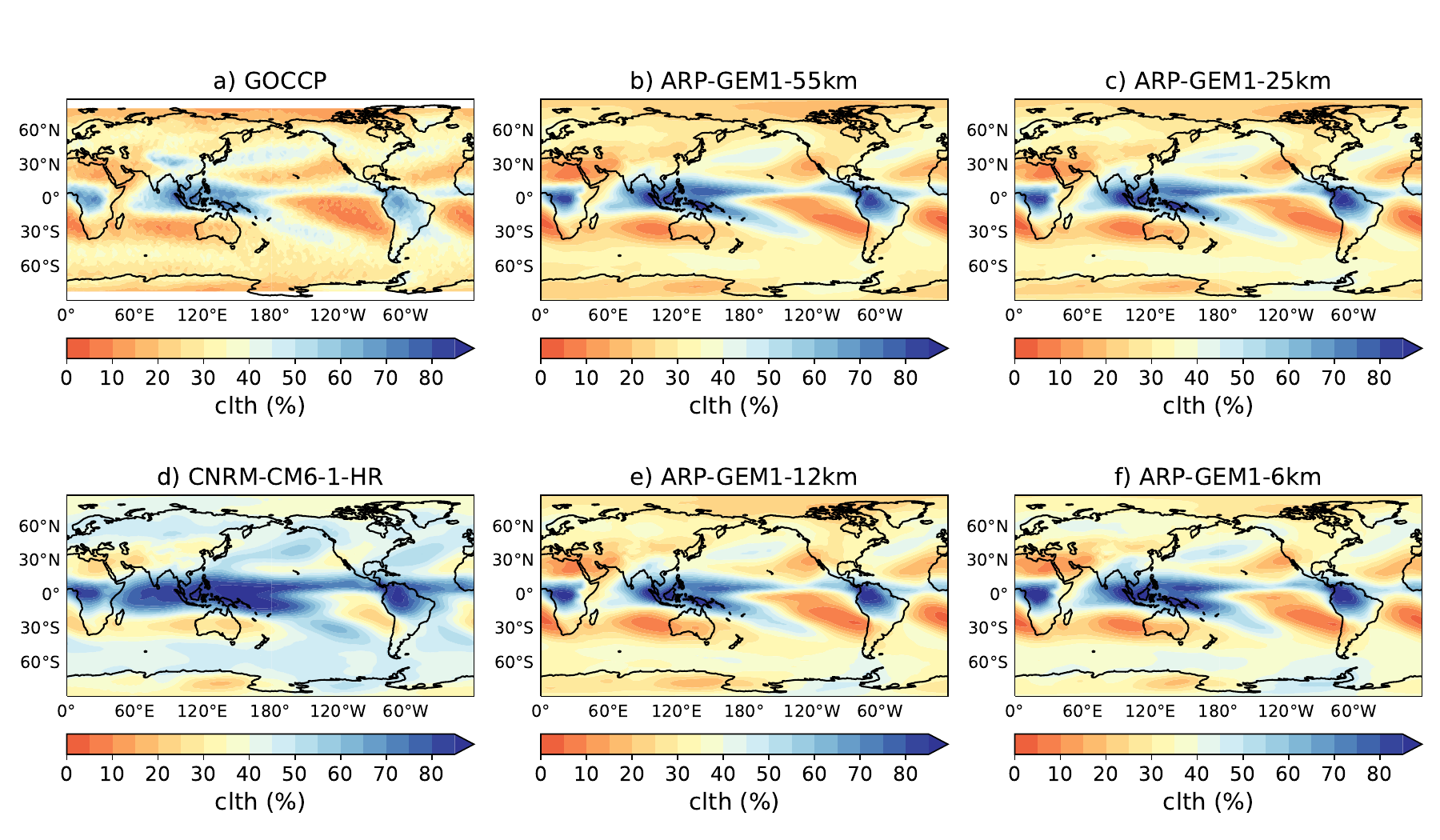}}
\caption{Annual mean of high-level cloud fraction (period: 2007-2012) for : (a) CALIPSO-GOCCP data, (b) ARP-GEM1-55km, (c) ARP-GEM1-25km, (d) CNRM-CM6-1-HR, (e) ARP-GEM1-12km, and (f) ARP-GEM1-6km.}
\label{fig:map_clth}
\end{figure*}

We conduct a standard evaluation of the temporal mean climate states for the four ARP-GEM configurations as well as the CMIP6 model CNRM-CM6-1-HR. The comparison with CNRM-CM6-1-HR serves as a baseline to assess improvements in mean state variables related to changes in the model's physics.

\begin{figure*}[h]
\centerline{\includegraphics[width=33pc]{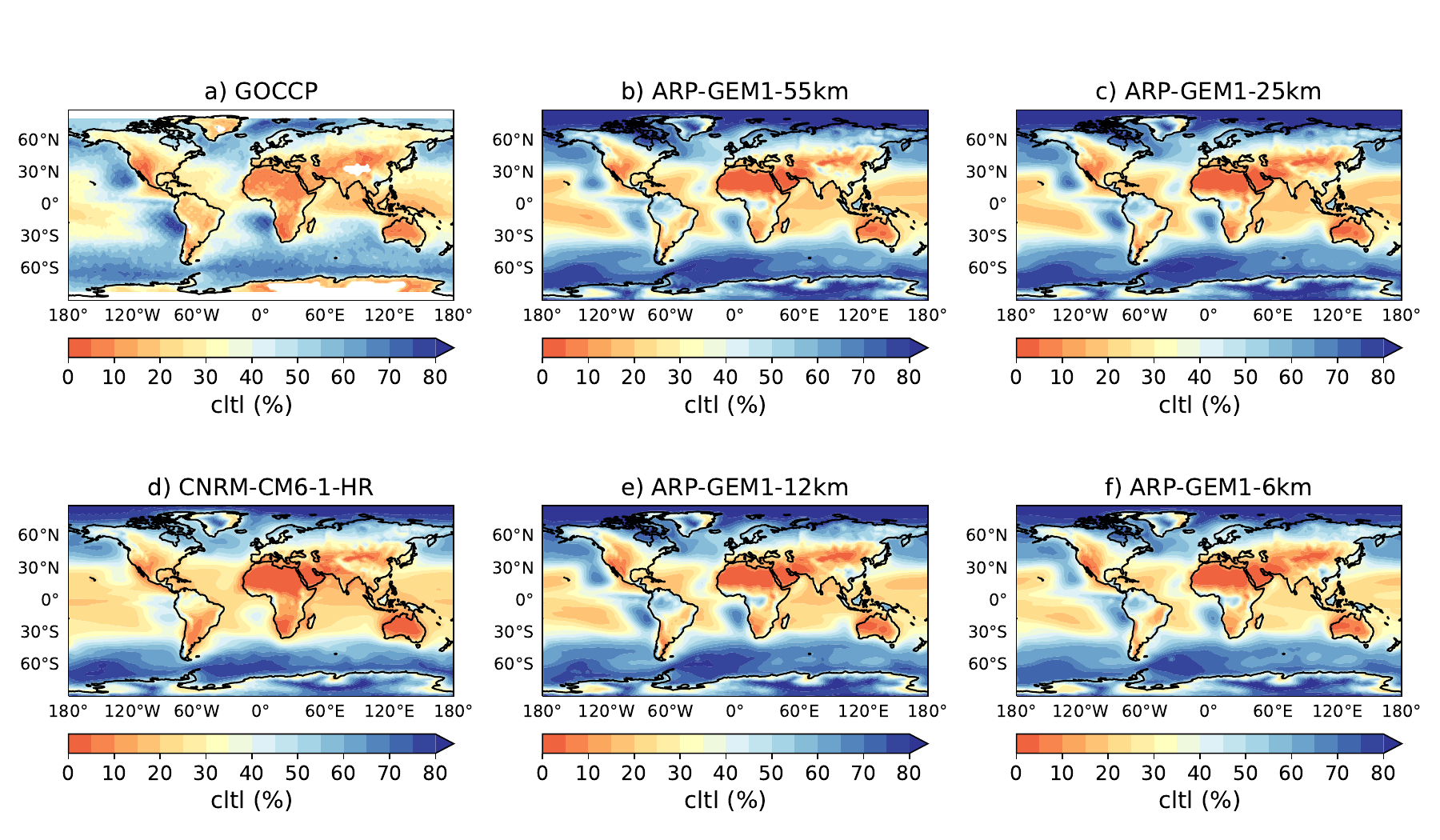}}
\caption{Same as Fig. \ref{fig:map_clth} for low-level cloud fraction.}
\label{fig:map_cltl}
\end{figure*}

Figures \ref{fig:map_rlut}-\ref{fig:map_cltl} display global maps of top-of-the-atmosphere atmospheric radiative variables, including outgoing LW radiation (OLR), the SW cloud radiative effect (CRE), and both low-level and high-level cloud cover. CRE is closely related to cloud biases. For LW fluxes, we focus on OLR rather than LW CRE, as the modelled LW CRE differs from observational estimates for methodological reasons \citep[e.g.,][]{loeb-2020}.

The global maps confirm the results presented in the previous section, showing small spatial biases for both LW and SW radiation, as well as low-level and high-level cloud covers for the ARP-GEM model. Improvements in both LW and SW radiation (Figs \ref{fig:map_rlut}-\ref{fig:map_rstcre}) are tied to better cloud distributions, including improvements in high cloud cover for both LW and SW radiation (Fig. \ref{fig:map_clth}b, c) and a more accurate spatial distribution of low-level clouds (Fig. \ref{fig:map_cltl}b, c) for SW radiation. These improvements help prevent excessive overfitting through parameter calibration, particularly by adjusting inhomogeneity factors.

Few biases are observed in the LW spatial pattern. The improvement in the representation of high clouds, including a reduction in cloud amount (Fig. \ref{fig:map_clth}), leads to better LW fluxes compared to CMIP6. Changes in cirrus clouds are related to the replacement of the cloud scheme and, to some extent, the change in the deep convection scheme, along with revised microphysics calibration, particularly regarding ice autoconversion.

\begin{figure}
\centerline{\includegraphics[width=17pc]{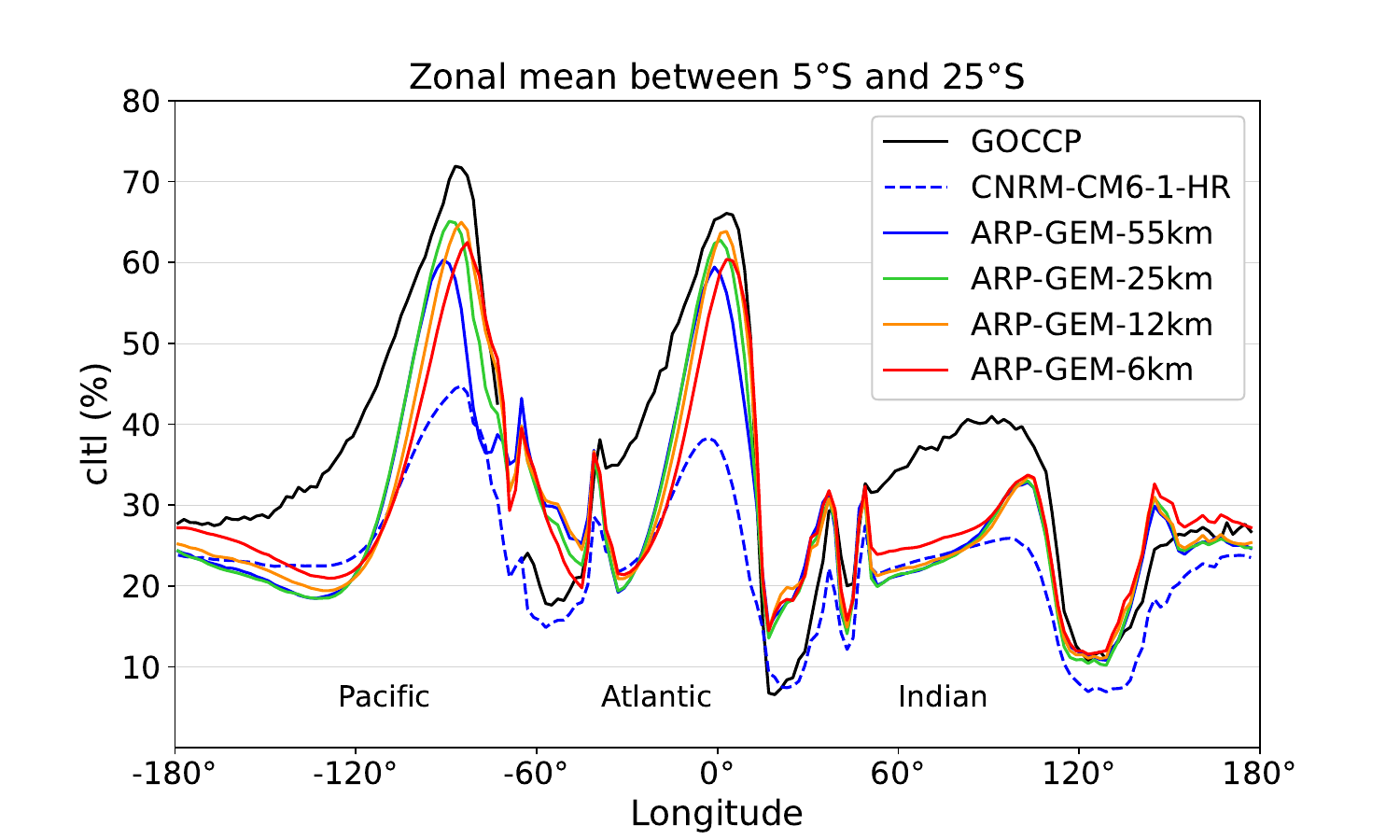}}
\caption{Zonal mean low-level cloud fraction in the latitude band 25$^\circ$S - 5$^\circ$S as a function of longitude.}
\label{fig:zonal_cltl}
\end{figure}

SW fluxes are better represented in ARP-GEM1-55km compared to CNRM-CM6-1-HR. However, limitations remain, with biases observed in the Southern Hemisphere -- a common issue in climate models \citep[e.g.,][]{grise-2014} -- and in the Eastern ocean basins. Nevertheless, the bimodal character of low-level clouds is better captured in ARP-GEM, with a larger cloud fraction in the Eastern oceans' stratocumulus regions (Fig. \ref{fig:map_cltl} and Fig. \ref{fig:zonal_cltl}). This suggests that some improvements are possible through tuning, preliminarily to a revision of the schemes. In the CMIP6 version, the lack of stratocumulus clouds was compensated by excessively large cloud fractions elsewhere, using a multiplication factor of 10 to derive cloud fraction from the updraft area. 

Improvements in stratocumulus are linked to modifications in the shallow convection scheme, particularly the management of overshoot, which is limited in cases of large inversion. The reduced shallow convection updraft prevents mixing at the top of the boundary layer, resulting in a more accurate depiction of cloud cover in stratocumulus regions, similar to the changes described in \citet{hourdin-2019}. To a lesser extent, the updated cloud scheme supports better calibration of low-level cloud cover. 

Additionally, ARP-GEM shows a better vertical structure of low-level clouds, with cloud tops rising progressively further offshore (not shown), akin to the behavior shown in the LMDZ6A climate model with similar revision in the shallow convection scheme \citep{hourdin-2019}. In contrast, CNRM-CM6-1-HR exhibits a constant cloud top altitude, as previously shown by \citet{brient-2019}. 

With increasing resolution, biases related to low-level clouds and the associated SW cloud radiative effect are reduced (Fig. \ref{fig:map_rstcre} and \ref{fig:map_clth}). In particular, the representation of marine stratocumulus clouds is improved near the coast with higher spatial resolution. This is evident in the longitudinal mean of cloud cover, which shows increased cloud cover closer to land (Fig. \ref{fig:zonal_cltl}). 
Additionally, SW biases are reduced in some areas, possibly due to effects related to topography.

An indirect factor may contribute to the observed improvements in shortwave radiation and low cloud amount. Specifically, the use of a lower critical relative humidity $RH_{\mathrm{c},\mathrm{low}}$ in higher resolution configurations (see Table \ref{tab:tuning}), aimed at compensating for the reduction in cloud amount, results in larger subgrid cloud distributions. This adjustment may increase the amount of low clouds across many regions. Thus, some improvements might be attributed to favorable tuning in this model configuration, rather than resolution alone.

Despite general improvements, biases persist, particularly in the transition from stratocumulus to cumulus clouds, which is a common issue in climate models. Further enhancements in physical parameterizations are required to better represent coastal regions and refine the stratocumulus-to-cumulus transition.
Note that LW radiation shows an increasing bias at the North Pole, which becomes particularly pronounced in the 6-km resolution (Fig. \ref{fig:map_rlut}). The reasons for this bias are unclear, with multiple factors potentially involved.

\subsection{Surface air temperature}
\label{sec:tas}

\begin{figure*}[h]
\centerline{\includegraphics[width=33pc]{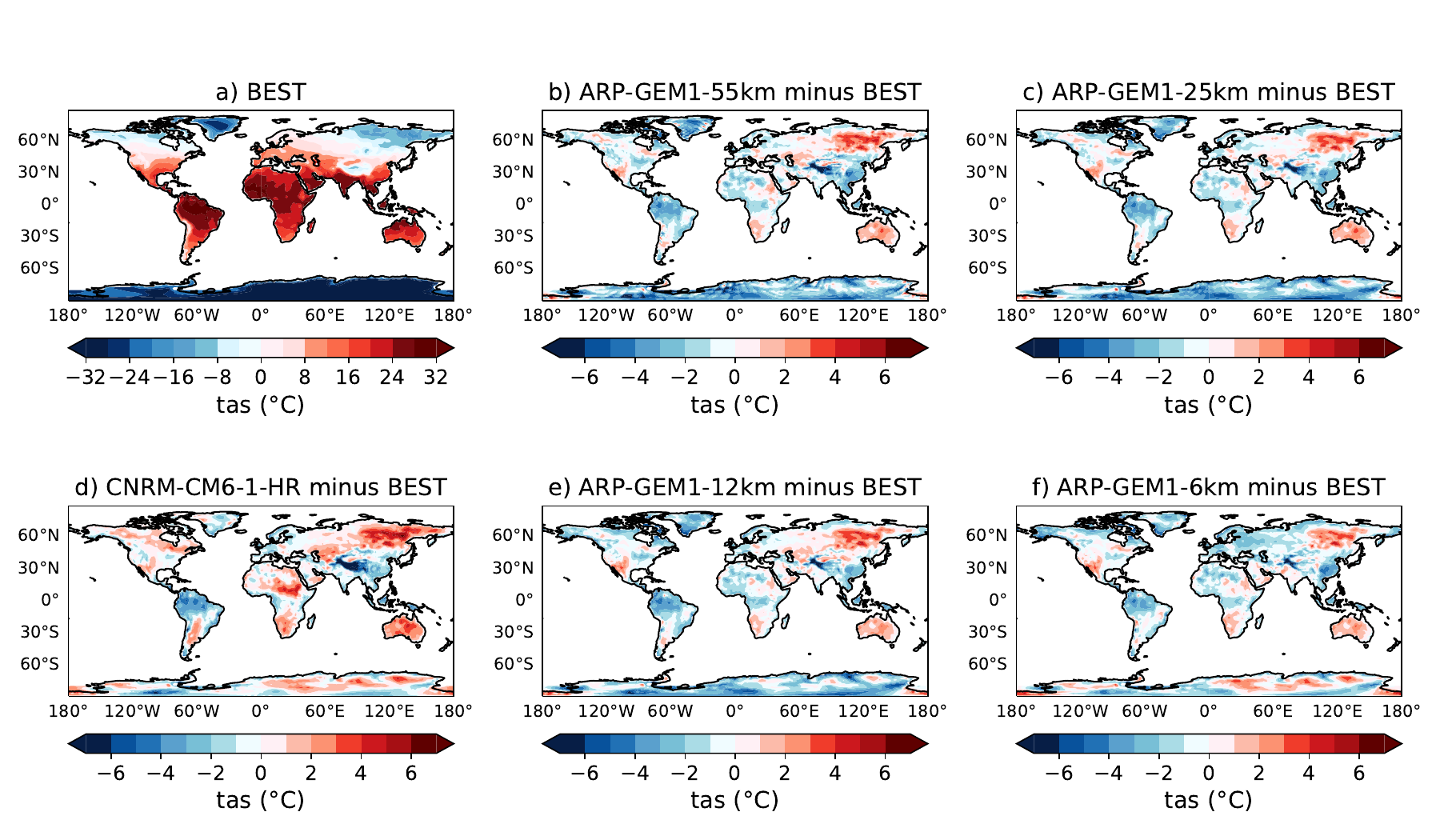}}
\caption{Same as Fig. \ref{fig:map_rlut} for surface air temperature, with BEST data as reference, over the period 1985-2014.}
\label{fig:map_tas}
\end{figure*}

\begin{figure*}[h]
\centerline{\includegraphics[width=33pc]{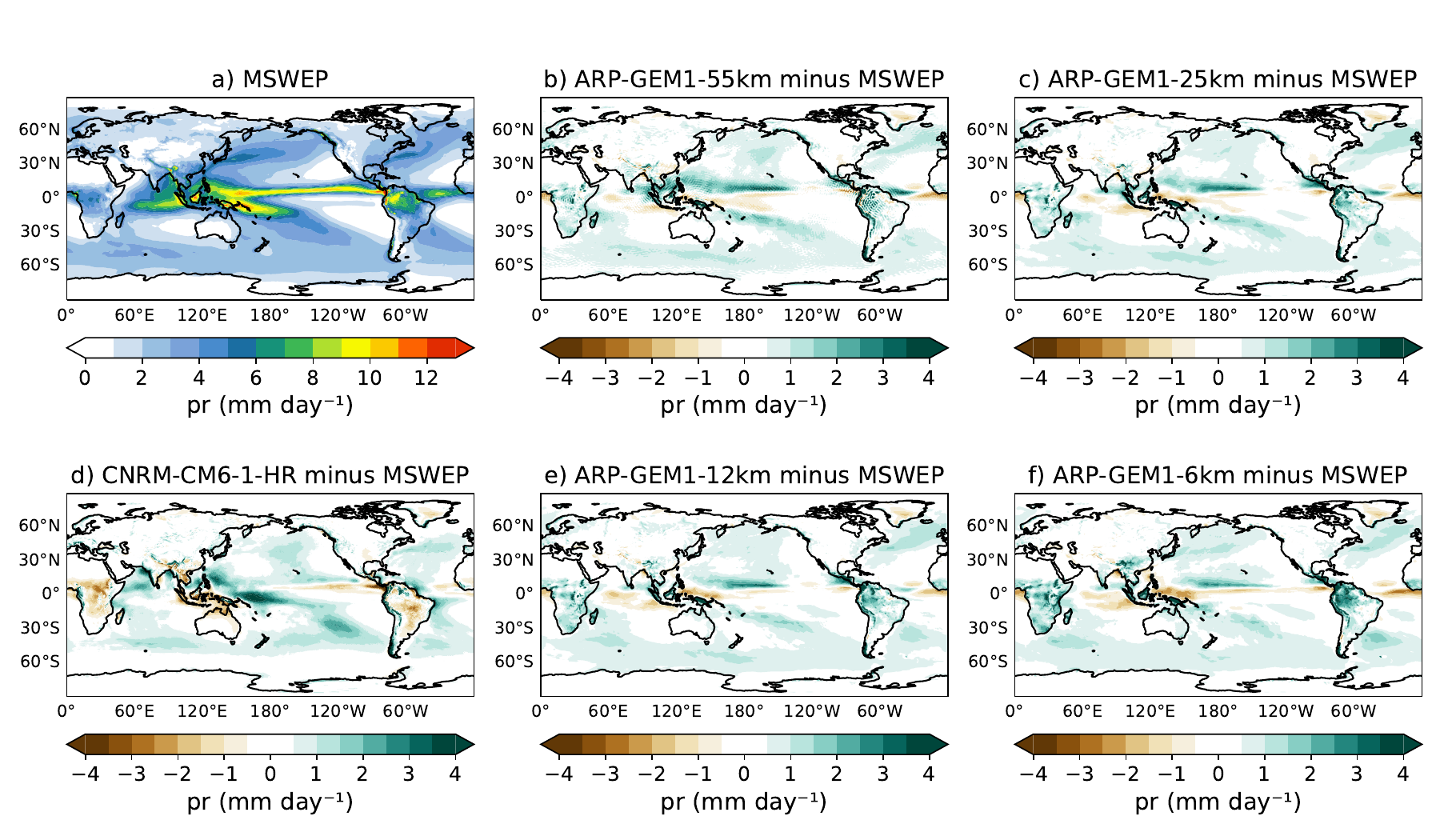}}
\caption{Same as Fig. \ref{fig:map_rlut} for precipitation, with MSWEP data as reference, over the period 1985-2014.}
\label{fig:map_pr}
\end{figure*}

The surface air temperature has improved in ARP-GEM1 compared to CNRM-CM6-1-HR. However, errors persist, and there is room for further improvement (Fig. \ref{fig:map_tas}). The limited representation of surface temperature is primarily due to a cold bias in most areas and a warm bias mainly in Northern Eurasia. This bias is particularly pronounced in winter (not shown), a season where errors are more significant across CMIP climate models (e.g. Fig. \ref{fig:rmse}b). Additionally, a large number of CMIP models exhibit a similar bias pattern to ARP-GEM1 \citep{fan-2020}, suggesting the presence of systematic errors across models.

The resolution generally improves the surface air temperature, though the cold bias persists. This may be attributed to a more accurate representation of surface elevation. Finally, reducing biases may be achieved through better representation of low-level mixing. Positive effects have been observed from tuning turbulence. Significant improvements will likely require revisions in shallow convection and turbulence, which are currently under investigation.

\subsection{Precipitation}
\label{sec:precip}

The precipitation pattern is well represented in ARP-GEM, despite a general overestimation of precipitation (Fig. \ref{fig:map_pr}). The CNRM-CM6-1-HR pattern, with a deficit of precipitation over ascending branch of the Walker circulation (land and maritime continent), is associated with relatively large entrainment values (to compensate for the use of a single updraft), a small threshold for the maximum convective area allowed (to mitigate numerical instabilities), and initial updrafts starting from only the lowest model level (in contrast to the Jakob and Siebesma trigger). The combination of these features may produce an insufficiently active convection, leading to the observed pattern. The use of the Tiedtke-Bechtold scheme in the ARP-GEM1-55km, particularly when combined with the shallow convection scheme, leads to a clear improvement. 

\begin{figure*}
\centerline{\includegraphics[width=33pc]{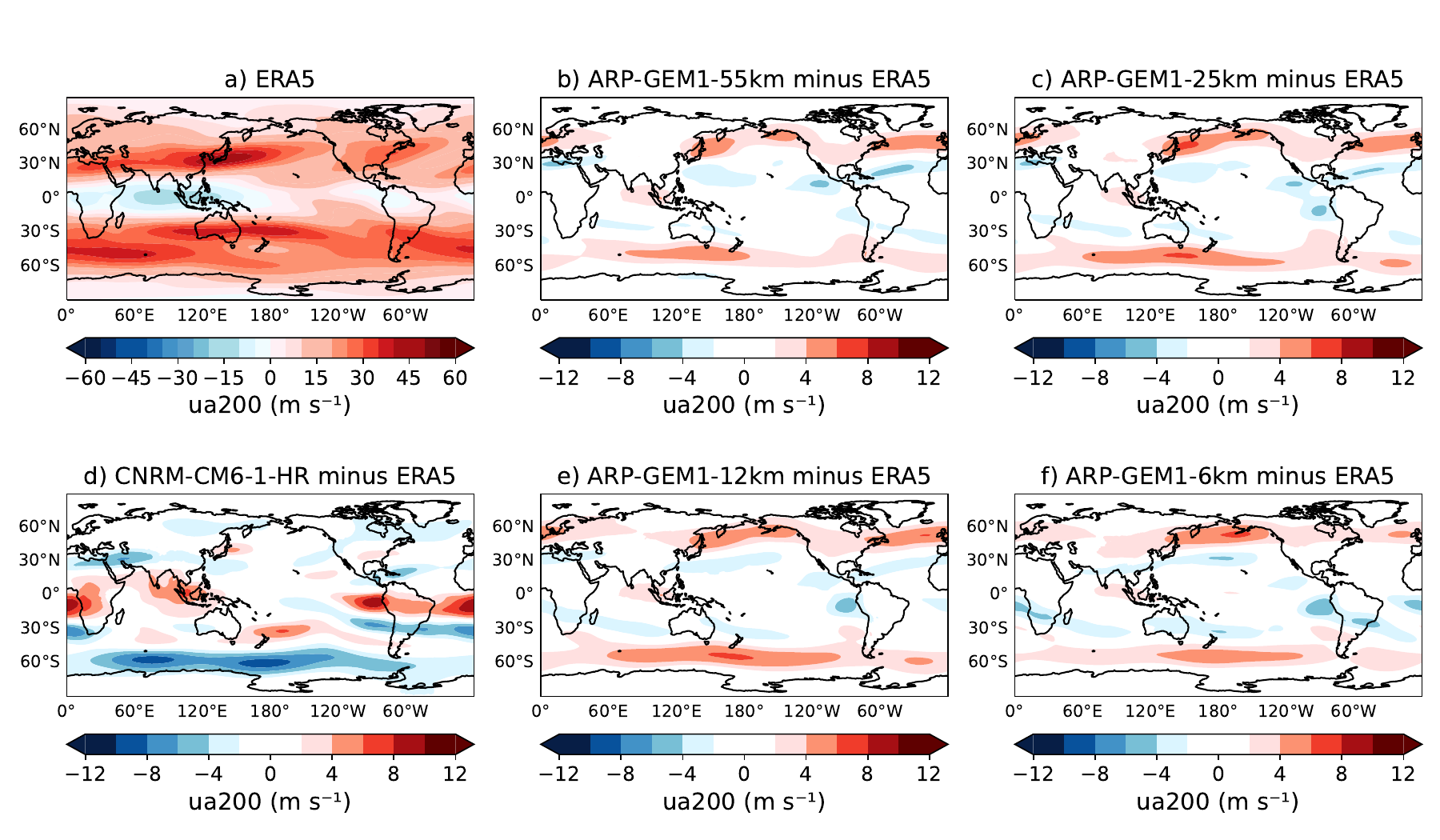}}
\caption{Same as Fig. \ref{fig:map_rlut} for 200 hPa zonal wind, with ERA-5 data as reference, over the period 1985-2014.}
\label{fig:map_ua200}
\end{figure*}

In the higher-resolution configurations, precipitation appears more dispersed, with fewer localized patches of heavy precipitation. While precipitation is best represented in the 12-km simulation, the 6-km simulation shows larger errors, with significant biases predominantly over land. Further refinement and tuning are required to address these discrepancies. Adjusting precipitation is challenging due to its sensitivity to a complex chain of processes and numerous interacting factors. Preliminary tests suggest that improving the representation of precipitation associated with shallow convection is a necessary step for better precipitation modeling in this model.

\subsection{Upper zonal wind and jets}

We present a global map of the 200-hPa zonal wind velocity (Fig. \ref{fig:map_ua200}), offering insight into the representation of subtropical jet streams. For comparison, we use ERA-5 reanalysis data \citep{hersbach-2020}. Overall, the wind is well represented, though the jet streams are positioned too far north in these simulations. In the CMIP6 version, there are pronounced biases in the tropics. Improvements in ARP-GEM1 are linked to the replacement of the convection scheme, particularly through the role of momentum transport by convection. In high-resolution configurations, the wind is better represented, likely influenced by the increased resolution of the topography.

\subsection{Tropospheric humidity and temperature}
\label{sub:ta_hus}

\begin{figure*}
\centerline{\includegraphics[width=35pc]{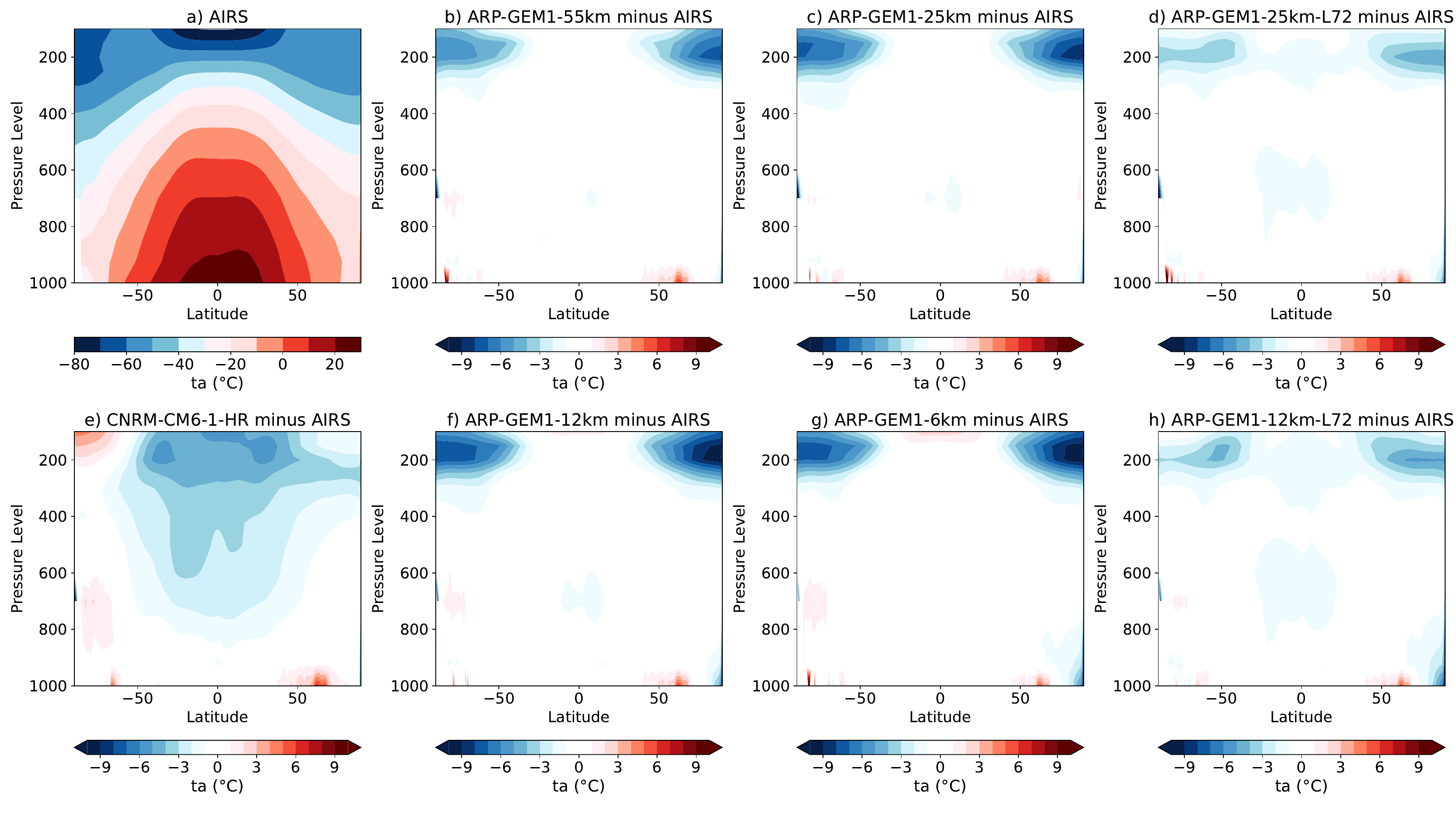}}
\caption{Annual and zonal mean of (a) air temperature climatology from AIRS (period 2003-2018) and (b-h) air temperature bias (period 2003-2014) for (b) ARP-GEM1-55km, (c) ARP-GEM1-25km, (e) CNRM-CM6-1-HR, (f) ARP-GEM1-12km, and (g) ARP-GEM1-6km, and for the (d) 25-km and (h) 12-km configurations with 72 vertical levels.}
\label{fig:zonal_ta}
\end{figure*}

\begin{figure*}
\centerline{\includegraphics[width=35pc]{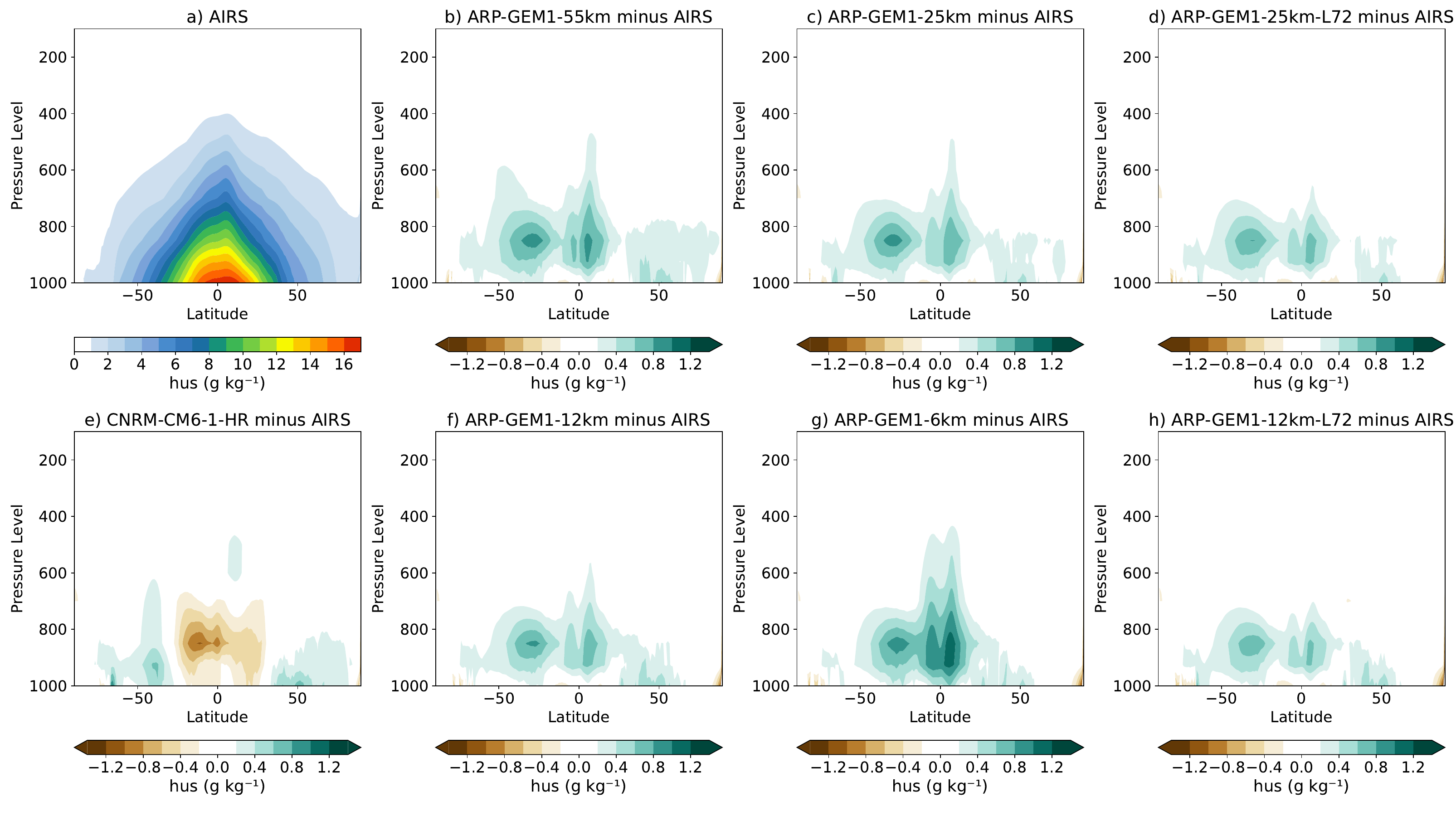}}
\caption{Same as Fig. \ref{fig:zonal_ta} for specific humidity.}
\label{fig:zonal_hus}
\end{figure*}

Figures \ref{fig:zonal_ta} and \ref{fig:zonal_hus} shows the zonal mean profiles of temperature and humidity. 
Observational data are from NASA's Atmospheric Infrared Sounder \cite[AIRS,][]{chahine-2006}. Both profiles are generally well-represented.
In CNRM-CM6-1-HR, the tropical upper troposphere was too cold, which is closely related to issues with the deep convection scheme (see Section \ref{sec:evaluation}\ref{sec:precip}). In contrast, ARP-GEM1 shows clear improvements in temperature representation, except at high latitudes where a cold bias has emerged at tropopause levels. In high-resolution configurations, this bias increases only slightly. This increase is expected due to the change in the grid’s aspect ratio (i.e., the ratio of horizontal to vertical resolution), but it is strongly mitigated by the use of the quintic interpolation \citep{polichtchouk-2020} in the semi-Lagrangian scheme (see Part I for details). 

Regarding humidity, the lower troposphere in CMIP6 was too dry, also reflecting the role of the deep convection scheme. In contrast, ARP-GEM1 configurations exhibit excessive moisture at lower levels, especially around the top of the boundary layer. Errors in low-level mixing associated with shallow convection likely contribute to this bias. Addressing other model biases, such as low-level cloud cover, often leads to an increase in this humidity bias. This indicates the need for further refinement in mixing parameterizations or an increase in  vertical resolution. 

\subsection{Sensitivity to vertical levels}

An additional set of experiments at 25-km and 12-km resolutions was conducted using a refined vertical grid with 72 vertical levels (Fig. 1 in Part I). As for other simulations, these simulations are radiatively equilibrated using the minimal tuning approach described in Section \ref{sec:simus}\ref{sec:tuning}. Figures \ref{fig:zonal_ta}d and \ref{fig:zonal_ta}h, displays the zonal mean profiles of temperature biases for both experiments. The temperature near the tropopause shows clear sensitivity to resolution, where large gradients necessitate fine resolution \citep[e.g.,][]{richter-2014}. A notable reduction in bias is achieved with this limited set of 72 vertical levels.

This refinement also contributes to a slight decrease low-level moisture bias, likely due to improved representation of boundary layer inversions (Figures \ref{fig:zonal_hus}d and \ref{fig:zonal_hus}h). Given the counteracting behavior of low-level humidity with other model biases, fewer vertical levels can enhance calibration and process selection. The 200 hPa wind is also improved in these simulations (not shown). Other variables are not significantly impacted by this refined grid. Although a more focused study would be needed to document clear sensitivities to resolution. The use of this grid will be considered for future ARP-GEM versions.

\subsection{Daily precipitation distribution}

\begin{figure}
\centerline{\includegraphics[width=19pc]{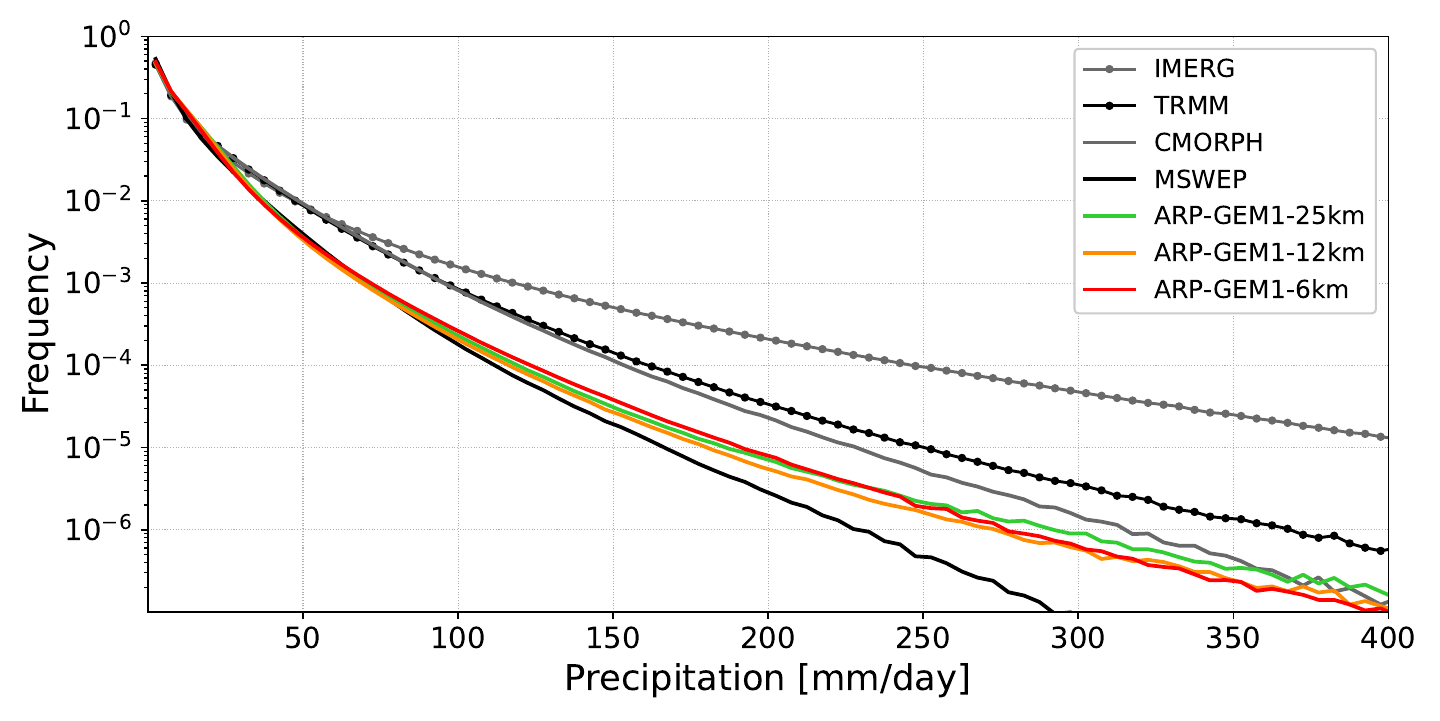}}
\caption{Probability density function of daily-mean precipitation (unit : mm/day) over tropical domain ($20^{\circ}$S-$20^{\circ}$N) for IMERG, CMORPH, TRMM and MSWEP datasets and for the ARP-GEM1 simulations with 55, 25, 12, and 6 km. The period used is 2001-2012 for all datasets. Precipitation is conservatively interpolated to a 0.25$^{\circ}$ $\times$ 0.25$^{\circ}$ grid (excepted for the 55 km). The bin size is 5 mm.day$^{-1}$.}
\label{fig:pdf_precip}
\end{figure}

Observed and modelled probability density functions (PDFs) of daily tropical ($20^{\circ}$S-$20^{\circ}$N) precipitation are shown in Figure \ref{fig:pdf_precip}. Four observational datasets are used : the Integrated Multi-satellitE Retrievals for the Global Precipitation Measurement (IMERG) dataset, version 06 \citep{huffman-2019}, the Tropical Rainfall Measuring Mission (TRMM) dataset, version 7\_3B42 \citep{huffman-2007}, the Climate Prediction Center MORPHing technique (CMORPH) product, version 1.0 \citep{xie-2017} and the Multi‐Source Weighted‐Ensemble Precipitation (MSWEP) dataset, version 1.2 \citep{beck-2017}. For intense precipitation events, large differences between these products exist \citep{roca-2021}.

The PDFs simulated by the ARP-GEM1 model are within the range of the observations used. For extreme precipitations (exceeding 250 mm/day), the ARP-GEM1 model closely aligns with the CMORPH and TRMM datasets but yields lower values compared to the IMERG product. The sensitivity to horizontal resolution is small, with the PDFs from all three configurations of ARP-GEM1 appearing very similar. These results suggest that the 6-km horizontal grid is not sufficient to significantly alter the representation of extreme precipitation events and that the resolved part of the convective motions (and their associated rain) remains small at the 6-km resolution.

\subsection{Tropical cyclones}

\begin{figure*}
\centerline{\includegraphics[width=30pc]{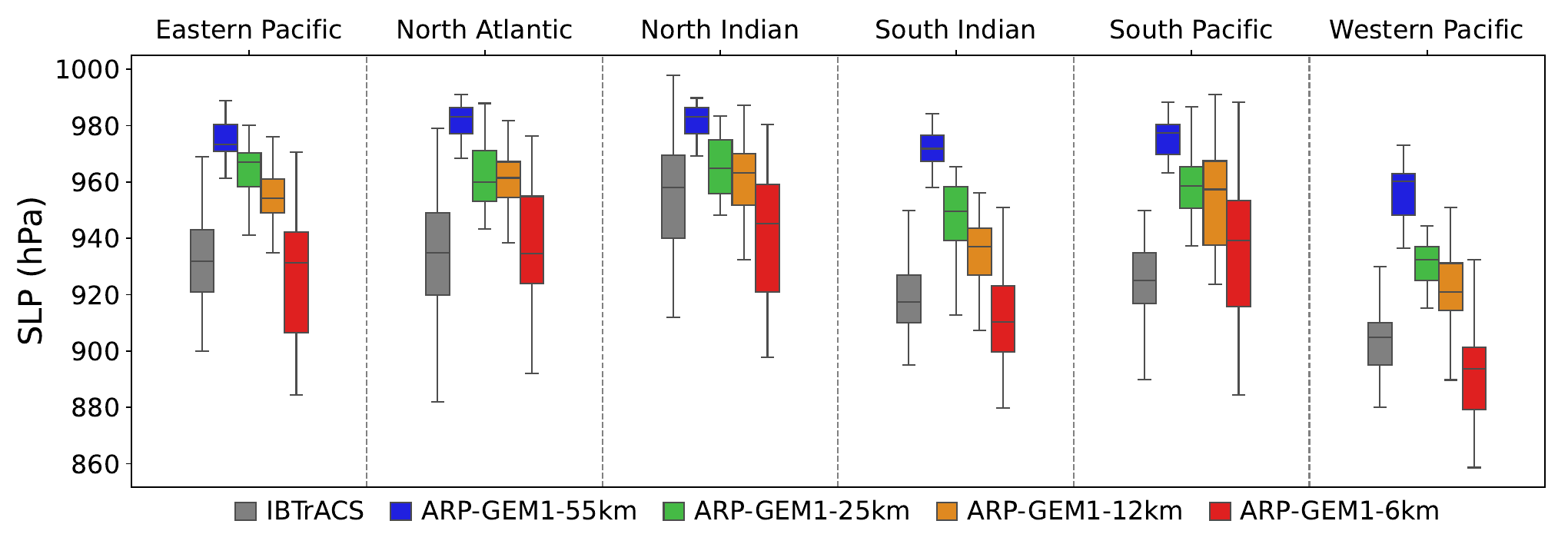}}
\caption{Distribution of the annual minimum value of 6-hourly sea-level pressure (hPa) for IBTrACS and ARP-GEM1 experiments at 55, 25, 12, and 6-km resolution (period : 1985-2014) across the six TC basins : Eastern Pacific (5$^{\circ}$N/35$^{\circ}$N, 180$^{\circ}$E/260$^{\circ}$E), North Atlantic (5$^{\circ}$N/35$^{\circ}$N, 260$^{\circ}$E/350$^{\circ}$E), North Indian (5$^{\circ}$N/30$^{\circ}$N, 45$^{\circ}$E/100$^{\circ}$E), South Indian (-30$^{\circ}$S/-5$^{\circ}$S, 30$^{\circ}$E/130$^{\circ}$E), South Pacific (-30$^{\circ}$S/-5$^{\circ}$S, 135$^{\circ}$E/230$^{\circ}$E) and Western Pacific (5$^{\circ}$N/35$^{\circ}$N, 105$^{\circ}$E/175$^{\circ}$E). In all boxplots, the box represents the first and third quartiles, the band inside is the median, the whiskers expand to the largest values still within the 1.5 interquartile range from the box.}
\label{fig:boxplot_TC}
\end{figure*}

To assess the realism of tropical cyclones (TCs) simulated by the different ARP-GEM configurations, we compare statistical distributions of the annual minimum sea-level pressure against observed data using a method proposed by \citet{cattiaux-2020}. Originally designed for a straightforward assessment of TCs in the Southern Indian Ocean, this method is applied here to evaluate TCs across all global cyclonic basins. The PDFs of the annual minimum sea-level pressure are computed from the model's pressure fields in each oceanic basin. These PDFs are compared with those derived from the IBTrACS dataset, which provides sea-level pressure data along cyclonic trajectories. 

Specifically, we use the IBTrACS version 4 dataset, released in April 2019, which provides track characteristics (e.g., position, sea-level pressure, maximum sustained winds) on a 3-hourly basis \citep{knapp-2010}. We restrict the IBTrACS data to the period from July 1979 to June 2016 (covering cyclone seasons from 1980 to 2016), focusing on hours 0000, 0600, 1200, and 1800 UTC, and include only systems labeled as 'TS' (tropical storm) that reached at least 10 m/s at some point in their lifetime. For each retained TC, we compute the minimum sea-level pressure and retain this minimum for each annual cyclonic season across all basins. For the ARP-GEM1 model versions, the annual minimum sea-level pressure is computed using the 6-hourly instantaneous sea-level pressure field for each cyclonic basin.

\begin{figure*}
\centerline{\includegraphics[width=33pc]{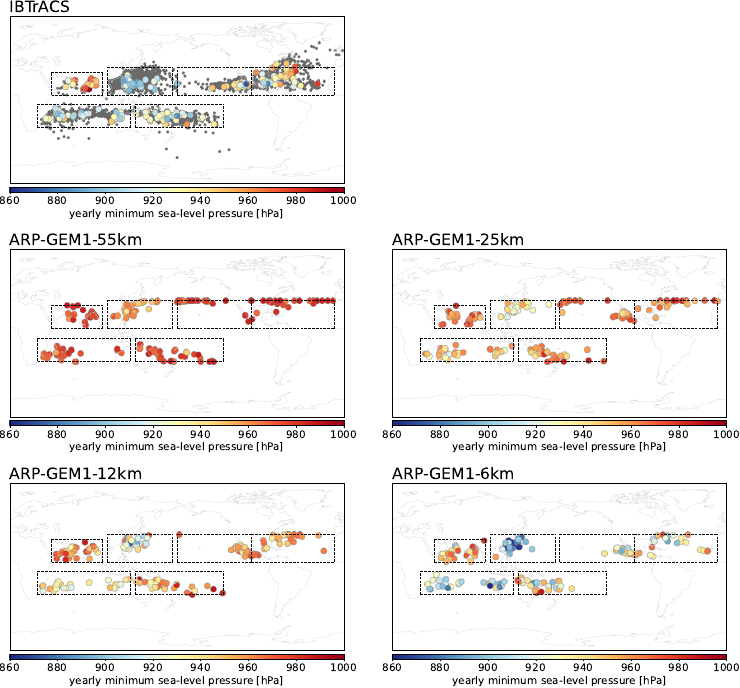}}
\caption{Annual minimum values of 6-hourly sea-level pressure for IBTrACS and ARP-GEM1 experiments at 55, 25, 12, and 6-km resolution (period : 1985-2014) across the six tropical cyclone basins : Eastern Pacific (5$^{\circ}$N/35$^{\circ}$N, 180$^{\circ}$E/260$^{\circ}$E), North Atlantic (5$^{\circ}$N/35$^{\circ}$N, 260$^{\circ}$E/350$^{\circ}$E), North Indian (5$^{\circ}$N/30$^{\circ}$N, 45$^{\circ}$E/100$^{\circ}$E), South Indian (-30$^{\circ}$S/-5$^{\circ}$S, 30$^{\circ}$E/130$^{\circ}$E), South Pacific (-30$^{\circ}$S/-5$^{\circ}$S, 135$^{\circ}$E/230$^{\circ}$E) and Western Pacific (5$^{\circ}$N/35$^{\circ}$N, 105$^{\circ}$E/175$^{\circ}$E). Gray dots correspond to the localization of the minimum values of sea-level pressure for all tropical cyclones selected in the IBTrACS dataset.}
\label{fig:map_TC}
\end{figure*}

\begin{figure*}
\centerline{\includegraphics[width=33pc]{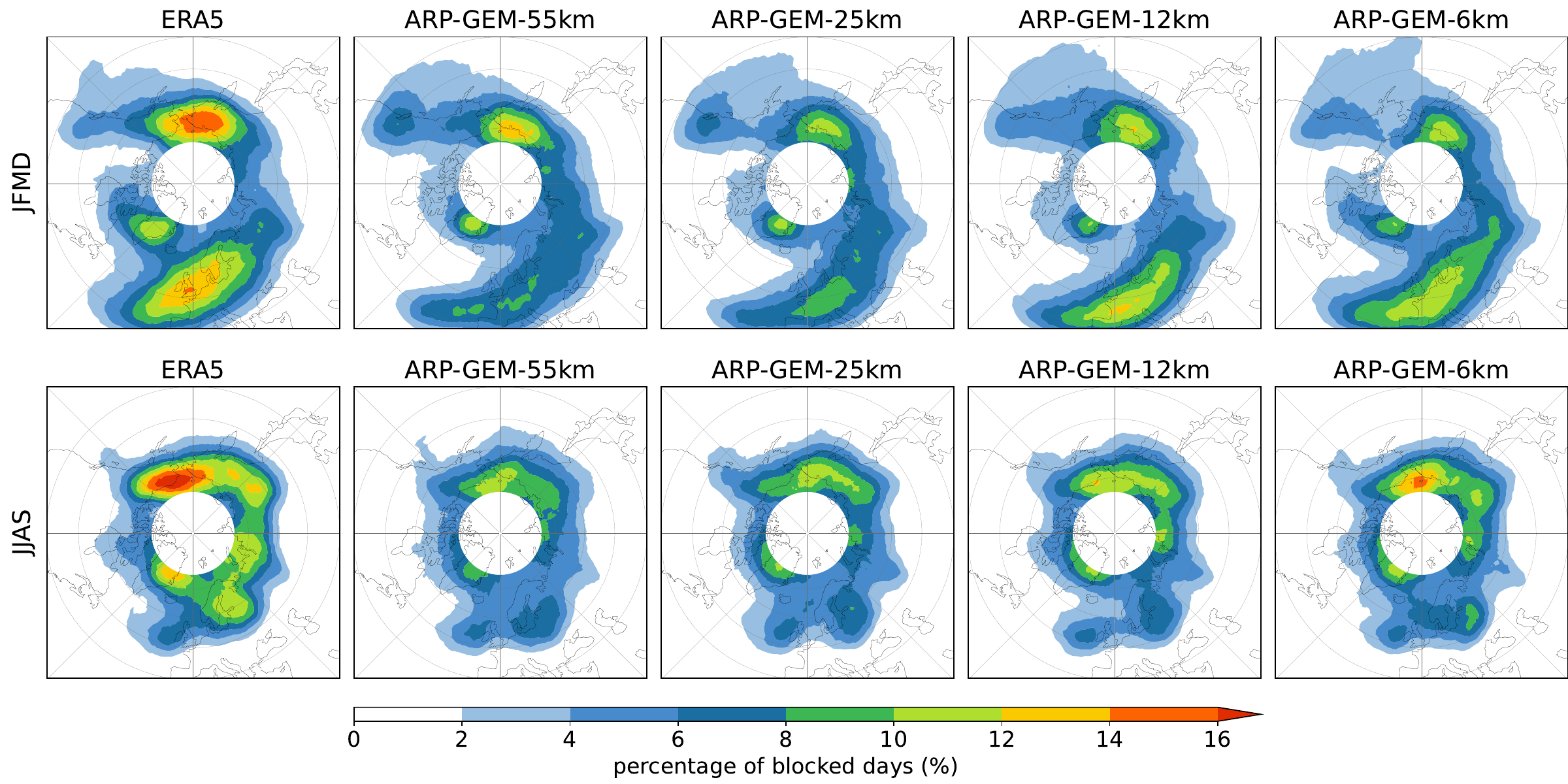}}
\caption{Northern Hemisphere blocking climatology (unit : percentage of blocked days) for ERA5 (period : 1985-2014) and for the ARP-GEM-1 models at 55, 25, 12 and 6 km (period : 1985-2014). Rows cover (top) winter (DJFM) and (bottom) summer (JJAS). All model and reanalysis fields (period: 185-2014) are preliminarily regridded to a common 1.25$^{\circ}$ $\times$ 1.25$^{\circ}$ regular grid before the blocking identification is applied.}
\label{fig:blocking}
\end{figure*}

The results are shown in Figure \ref{fig:boxplot_TC}, where the multi-annual distributions are presented as boxplots. In the lower resolution runs (55 km and 25 km), the minimum pressures are clearly too high compared to those observed along cyclonic trajectories. Both the 12-km and 6-km resolutions yield satisfactory results, particularly the 6-km resolution simulation, for which the probability density function (PDF) closely matches the observed data, highlighting the added value of higher resolution in representing cyclones.

Figure \ref{fig:map_TC} shows the spatial locations of each pressure minimum in each ocean basin and compares them to the cyclonic pressure minima in the IBTrACS dataset. At lower resolutions, the minima are not necessarily located in cyclonic areas, suggesting that they may be associated with storms rather than true cyclones. This reinforces the conclusion that the model is unable to effectively capture cyclones at lower resolutions. In contrast, at the 12 km and especially at the 6 km resolutions, the minima are correctly positioned within cyclonic regions.

\subsection{Northern Hemisphere blocking occurences}

The representation of midlatitude variability is assessed by examining the occurrence of atmospheric blocking events, which are quasi-stationary high-pressure systems that persist for several days and alter the midlatitude westerly flow. Preferred regions for the occurrence of blocking are the eastern sides of the Pacific and Atlantic Oceans. We follow the blocking identification method of \citet{davini-2012}, which is an extension of the blocking index described by \citet{tibaldi-1990}.  The blocking occurrences are computed from daily average geopotential height fields for all Northern Hemisphere grid points.

Figure \ref{fig:blocking} shows the blocking frequency for the different resolutions of ARP-GEM-1 in winter (top row) and summer (bottom row). The results are compared with the reference reanalysis ERA-5 field. Very similar results are obtained for ERA-I and JRA-55 reanalysis products. All resolutions capture the hemispheric-scale pattern of blocking frequency maxima in the Euro-Atlantic and Pacific regions but tend to underestimate the blocking frequencies. Increasing the resolution improves the representation of the blocking occurrence in the Euro-Atlantic areas but not in the Pacific Ocean. The number of blocking events is reduced over Greenland. These results are consistent with those obtained in \citet{schiemann-2017}, which compares the resolution sensitivity of North Hemisphere blocking in four atmospheric models. For the summer season, blocking frequencies are underestimated by the ARP-GEM1 model at different resolutions. This is particularly true in the Baltic region. The blocking occurrence in the Pacific basin is better captured by the 6-km version of the ARP-GEM1 model.

\section{Conclusion}

In this series of two articles, we describe and evaluate the first version of the ARP-GEM global atmospheric model. In this second part, we present a suite of simulations at resolutions ranging from 55 km to 6.3 km. The model demonstrates high computational efficiency at lower resolution (as discussed in Part I), and this efficiency is maintained at the highest resolutions, achieving near-scalable performance.
Alongside its computational performance, the ARP-GEM model demonstrates reliability in representing global climate metrics, comparable to the best-performing CMIP6 models. These results confirm the relevance of the design choices made for ARP-GEM, and suggest that some of the acceleration strategies introduced in Part I could be effectively applied to other global atmospheric models.

Cloud cover is well represented and linked to various updates in the model physics. Improvements in low cloud cover primarily result from refinements in the shallow convection scheme. The use of a simpler cloud scheme has allowed for more precise tuning of cloud distributions. The deep convection scheme plays a significant role in improving the model, including its representation of precipitation, tropospheric winds, temperature, humidity, and cloud cover.

The ARP-GEM1 model still has room for improvement. The coherence between boundary layer parameterizations could be improved, particularly in the representation of cloud-top entrainment and the interactions between shallow thermals and turbulence. 
Moreover the intensity of shallow convection may be too strong over land at high resolution.
The microphysics scheme could benefit from more sophisticated representations or the introduction of optional schemes that may help address uncertainties. In present configurations, the stratosphere has crude representation that limits accurate stratosphere-troposphere interaction. This can be addressed by adding layers between 10 and 50 km and activating the non orographic gravity wave drag parameterization.

All simulations, whether high or low resolution, effectively represent the mean state while maintaining close radiation equilibrium, thereby validating the minimal tuning strategy. However, some differences in model performance do not scale directly with spatial resolution. The minimal parameter adjustments made during tuning are insufficient to address these discrepancies. A more systematic approach involving back-and-forth adjustments between configurations and with a larger set of parameters is needed to optimize performance across resolutions. New biases emerging at higher resolutions suggest that the model’s mean state is affected differently as resolution approaches the convection gray zone.  While these changes require further tuning or physics development, they may also reveal new aspects of the model's behavior, offering potential for future improvements.

The results presented here suggest that climate simulations can now be run at a 10-km scale with reasonable computational cost, even without accounting for performance improvements from newer computing architectures.  
This resolution is comparable to that used for many regional downscaling applications \citep[CORDEX, ][]{gutowski-2016}. At a global scale, this resolution offers several benefits, including a more accurate representation of phenomena such as cyclones. Additionally, it enhances fine-scale features such as marine low cloud cover near the coast and improves aspects related to better topographic resolution.

These simulations can serve as inputs for regional models run at the kilometer scale \citep{ban-2021}. Using a O(10)-km global model avoids the need for a configuration with three nested models: a global O(100)-km model forcing, a regional O(10)-km model, which in turn forces a local O(1)-km model. This approach helps reduce development efforts and minimizes errors due to regional modelling : inconsistencies arising from the discrepancies between the forcing model and the forced model \citep[e.g.,][]{moon-2024}. Moreover, given that regional simulations depend mostly on the accuracy of the driving model, ensuring the reliability of the global model is essential. 
For a 20-year period (1995-2014) of the 12-km paired simulations (\textit{amip} and \textit{amip-future4K}), the complete state of the model has been saved at 6-hr frequency, providing lateral boundaries conditions for regional convection permitting simulation with the model AROME \citep{seity-2011}.

Finally, a major challenge for climate models remains the representation of moist convection. Explicit resolution of deep convection surely requires O(1)-km simulations. These simulations remains very costly but results presented here suggest that pluriannual O(1)-km simulations will be available in a near future. The O(10)-km simulations presented here are valuable for preparing higher-resolution runs, identifying and addressing technical and physical issues, developing and testing new physical parameterizations and tuning strategies. 
More broadly, they represent a new valuable branch within the large family of models used to understand and simulate atmospheric processes at the climate scale.

\acknowledgments

\datastatement
For the ARP-GEM1 model outputs, please contact the authors. All CMIP6 model outputs, including CNRM-CM6-1-HR, are available via the portal: https://esgf-node.llnl.gov/search/cmip6. CERES data were obtained from https://ceres.larc.nasa.gov, CALIPSO/GOCCP from https://climserv.ipsl.polytechnique.fr/cfmip-obs, and the MSWEP dataset from www.gloh2o.org. TRMM 3B42 Version 7 datasets and AIRS data were obtained from https://disc.gsfc.nasa.gov. BEST temperature data are available at https://berkeleyearth.org/data, ERA5 data from the Copernicus Climate Data Store (https://cds.climate.copernicus.eu/), and the IBTrACS dataset from https://www.ncei.noaa.gov/products/international-best-track-archive.

\appendix[] 
\appendixtitle{List of models}
\label{sec:appendix_models}
The 38 model versions used in Section \ref{sec:evaluation}\ref{res:rmse} are ACCESS-CM2, ACCESS-ESM1-5, BCC-ESM1, CAMS-CSM1-0, CNRM-CM6-1, CNRM-CM6-1-HR, CNRM-ESM2-1, FGOALS-f3-L, GISS-E2-2-G, HadGEM3-GC31-LL, HadGEM3-GC31-MM, INM-CM4-8, INM-CM5-0, KACE-1-0-G, MIROC6, MIROC-ES2L, MPI-ESM1-2-HR, MRI-ESM2-0, NESM3, NorESM2-LM, SAM0-UNICON, UKESM1-0-LL, CESM2, CESM2-FV2, CESM2-WACCM, CMCC-CM2-SR5, CanESM5, E3SM-1-0, EC-Earth3, GFDL-CM4, GISS-E2-1-G, IITM-ESM, IPSL-CM6A-LR, IPSL-CM6A-MR1, MPI-ESM1-2-LR, MPI-ESM-1-2-HAM, NorCPM1, TaiESM1

\bibliographystyle{ametsocV6}

\end{document}